\documentclass[11pt,a4paper,useAMS,usenatbib]{emulateapj}
\usepackage{epsfig}
\usepackage{amsmath}
\usepackage{natbib}

\begin{document}

\title{\textit{Chandra} observations of NGC4342, \\ an optically faint, X-ray gas-rich early-type galaxy}

\author{\'Akos Bogd\'an\altaffilmark{1}, William R. Forman\altaffilmark{1}, Ralph P. Kraft\altaffilmark{1}, Christine Jones\altaffilmark{1}, Christina Blom\altaffilmark{2}, Scott W. Randall\altaffilmark{1}, Zhongli Zhang\altaffilmark{3}, Irina Zhuravleva\altaffilmark{3}, Eugene Churazov\altaffilmark{3}, Zhiyuan Li\altaffilmark{1}, Paul E. J. Nulsen\altaffilmark{1}, Alexey Vikhlinin\altaffilmark{1}, and Sabine Schindler\altaffilmark{4}}
\affil{\altaffilmark{1}Smithsonian Astrophysical Observatory, 60 Garden Street, Cambridge, MA 02138, USA}
\affil{\altaffilmark{2}Centre for Astrophysics \& Supercomputing, Swinburne University, Hawthorn, VIC 3122, Australia}
\affil{\altaffilmark{3}Max-Planck-Institut f\"ur Astrophysik, Karl-Schwarzschild-str. 1, 85748 Garching, Germany}
\affil{\altaffilmark{4}Institut f\"ur Astrophysik, Leopold-Franzens Universit\"at Innsbruck, Technikerstrasse 25, 6020
Innsbruck, Austria}
\email{E-mail: abogdan@cfa.harvard.edu}

\shorttitle{\textit{CHANDRA} OBSERVATIONS OF NGC4342}
\shortauthors{BOGD\'AN ET AL.}

\begin{abstract}
\textit{Chandra} X-ray observations of NGC4342, a low stellar mass  ($M_{\rm{K}}=-22.79$ mag) early-type galaxy, show luminous, diffuse X-ray emission originating from hot gas with temperature of $kT\sim0.6$ keV. The observed $0.5-2$ keV band luminosity of the diffuse X-ray emission within the $D_{25}$ ellipse is $L_{\rm{0.5-2keV}} = 2.7\times10^{39} \ \rm{erg \ s^{-1}}$. The hot gas has a significantly broader distribution than the stellar light, and shows strong hydrodynamic disturbances with a sharp surface brightness edge   to the northeast and a trailing tail. We identify the edge as a cold front and conclude that the distorted morphology of the hot gas is produced by ram pressure as NGC4342 moves through external gas. From the thermal pressure ratios inside and outside the cold front, we estimate the velocity of NGC4342 and find that it moves supersonically ($M\sim2.6$) towards the northeast. Outside the optical extent of the galaxy we detect $\sim$$17$ bright ($L_{\rm{0.5-8keV}} \gtrsim3\times10^{37} \ \rm{erg \ s^{-1}}$) excess X-ray point sources. The excess sources are presumably low-mass X-ray binaries (LMXBs) located in metal-poor globular clusters (GCs) in the extended dark matter halo of NGC4342. Based on the number of excess sources and the average frequency of bright LMXBs in GCs, we estimate that NGC4342 may host roughly $850-1700$ GCs. In good agreement with this, optical observations hint that NGC4342 may harbor $1200\pm500$ GCs. This number corresponds to a GC specific frequency of $S_{\rm{N}} = 19.9\pm8.3$, which is among the largest values observed in full-size galaxies. 
\end{abstract}

\keywords{galaxies: individual (NGC4342) --- galaxies: stellar content --- intergalactic medium --- X-rays: galaxies --- X-rays: ISM}

\section{Introduction}
The observed X-ray emission from low-stellar mass ($M<10^{11} \ \rm{M_{\odot}}$) early-type galaxies is usually determined by the population of LMXBs, whose typical luminosity is in the range of $10^{35}-10^{39} \ \rm{erg \ s^{-1}}$ \citep{gilfanov04}. Additionally, diffuse emission is also present, whose origin can be twofold. On the one hand, fraction of the (or up to the entire)  diffuse emission originates from faint ($10^{27}-10^{35} \ \rm{erg \ s^{-1}}$) unresolved compact objects, mostly  coronally active binaries (ABs) and cataclysmic variables (CVs) \citep{revnivtsev06,sazonov06}. On the other hand, truly diffuse X-ray emission may also be present from  sub-keV temperature ionized interstellar medium \citep[e.g.][]{mathews03,david06,bogdan11}. 

In the present paper we investigate the X-ray emitting components of NGC4342, a low-stellar mass lenticular (S0) galaxy \citep{de91} with an $\sim$$8$ Gyrs old stellar population. The major motivations of this study are (1) to understand the nature and physical properties of the diffuse X-ray emission in and around NGC4342 and (2) to explore the population of resolved sources. To carry out the analysis we rely on deep \textit{Chandra} X-ray observations of NGC4342. The sub-arcsecond angular resolution of \textit{Chandra}  permits us to resolve bright LMXBs even in the central regions of the galaxy with high source density.  Therefore the diffuse emission can be separated from the population of LMXBs, allowing us to perform in-depth studies of the various X-ray emitting components. 

 A peculiar property of NGC4342 is that it hosts an unusually massive black hole relative to its low bulge mass \citep{cretton99}, yielding one of the highest black hole-to-bulge mass ratios in the local Universe \citep{haring04}.  Recently, we studied the origin of this discrepancy between the massive black hole and the low bulge mass in NGC4342 \citep{bogdan12}. We demonstrated that the galaxy has a massive dark matter halo extending well beyond the optical size of the galaxy. The existence of a dark matter halo led us to conclude that the high black hole-to-bulge mass ratio is not the result of tidal stripping, in which the dominant fraction of the stellar population of NGC4342 was removed. Instead, NGC4342 formed as a low stellar mass system with a massive black hole, implying that the black hole and bulge did not grow in tandem \citep{bogdan12}.

\begin{figure}[t]
    \leavevmode\epsfxsize=8.5cm\epsfbox{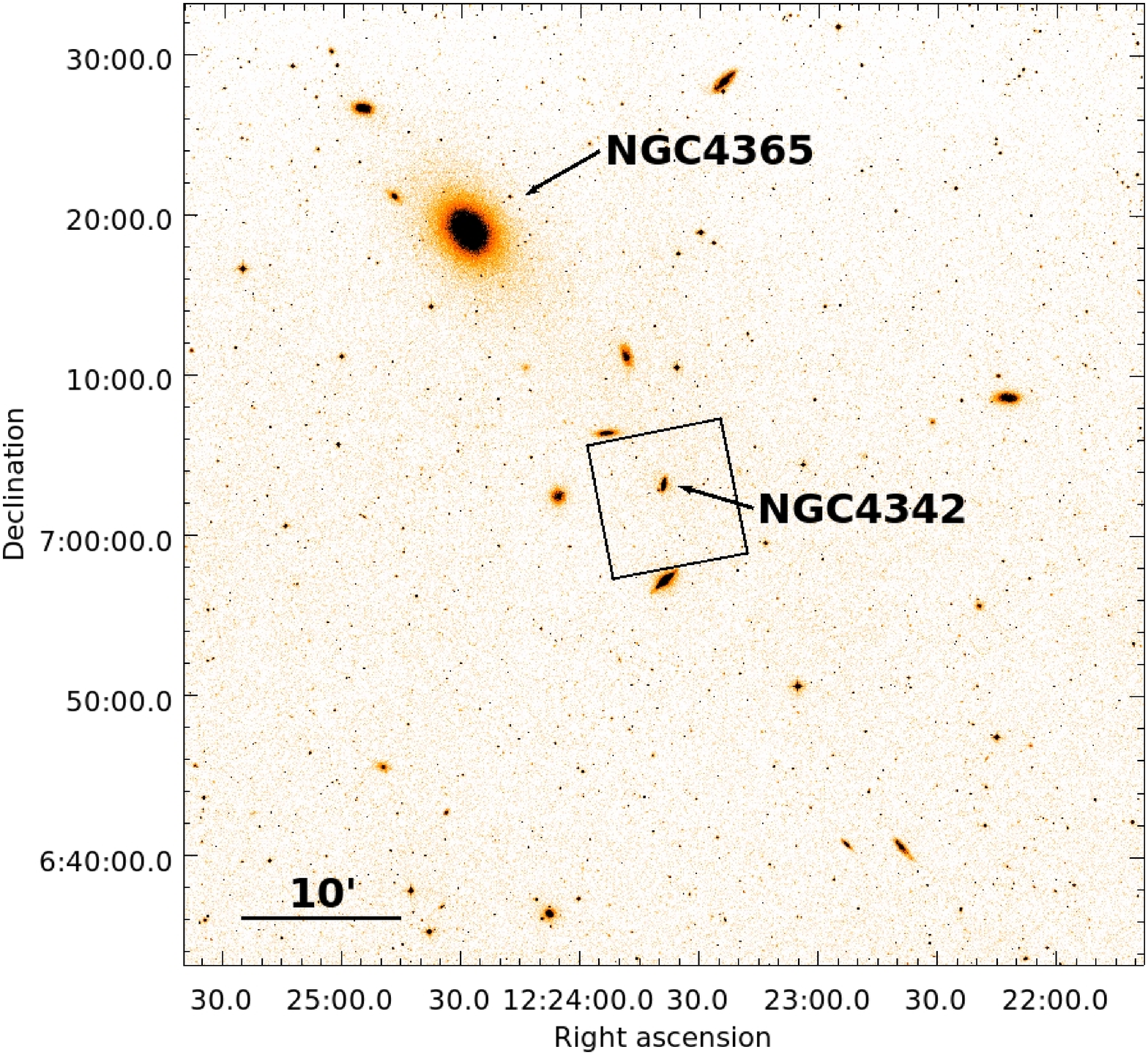}
    \caption{DSS R-band image of a $1\degr\times1\degr$ region around NGC4342, which is centered in the image. The massive elliptical galaxy, NGC4365 is located approximately $20\arcmin$ ($130$ kpc in projection) to the northeast of NGC4342. The neighborhood of NGC4342 is very dense, an overdensity of low mass galaxies lie in within  $0.5\degr$ ($200$ kpc in projection) of NGC4342. The overplotted box shows the field-of-view of the ACIS-S3 CCD.}
\label{fig:dss_big}
\end{figure}

In projection NGC4342 is located in the outskirts of the Virgo cluster,  $5.25\degr$ southwest from M87, the center of the Virgo cluster. The large-scale DSS R-band image, shown in Fig. \ref{fig:dss_big}, reveals a large galaxy density in the vicinity of NGC4342. Besides a number of low-mass galaxies, the massive elliptical galaxy, NGC4365, is located $20\arcmin$ to the northeast of NGC4342. The large galaxy density and the presence of substructure around  NGC4365 has already been recognized by \citet{de61}, who designated this substructure as the $W'$ cloud. In his morphological and kinematical study of the Virgo cluster, \citet{binggeli87} raised and studied the possibility that the $W'$ cloud is not part of the Virgo cluster, but is a galaxy group behind the cluster. In a more recent study, \citet{mei07} measured the distances of several galaxies associated with the $W'$ cloud using the surface brightness fluctuation method, and concluded that those galaxies lie at a distance of $23\pm1$ Mpc. In particular, \citet{mei07} found that NGC4365, which is presumably the most massive galaxy and hence the center of the group, is located at a distance of $23.3\pm0.3$ Mpc. Thus, the $W'$ cloud while projected on the Virgo cluster, but lies $\sim$$7$ Mpc behind it at a distance of $\sim$$23$ Mpc. Although NGC4365 is a relatively X-ray faint giant elliptical \citep{sivakoff03,bogdan11}, a recent study by \citet{blom12} demonstrated that its GC population extends well beyond the optical extent of the galaxy.  In particular, the population of blue GCs, which are believed to trace the dark matter halos of galaxies \citep[e.g.][]{bassino06,brodie06}, extend out to at least $134$ kpc \citep{blom12}. This indicates that NGC4365 presumably resides in a massive dark matter halo, hence it is feasible that the galaxy is the center of the $W'$ cloud.  Moreover, in Sect. \ref{sec:edge} we discuss that NGC4342 is   falling with supersonic velocity towards NGC4365, hence NGC4342 is also likely to belong the $W'$ cloud. Therefore we assume in the further discussion that NGC4342 is located at $D=23$ Mpc. At this distance  $1\arcsec$ corresponds to $111$ pc. 

Although NGC4342 most likely belongs to the $W'$ cloud behind Virgo cluster,  in the absence of accurate distance measurements, we cannot exclude the possibility that it is part of the NGC4472 sub-group of the Virgo cluster at a distance of $\sim$$16$ Mpc. NGC4472, the dominant member of the sub-group, is located at $\approx$$1.8^{\circ}$ towards the northeast of NGC4342, which corresponds to a projected distance of $\sim$$500$ kpc. We stress that while varying the distance of NGC4342 from $23$ Mpc (if associated with the $W'$ cloud) to $16$ Mpc (if associated with the Virgo cluster) changes some of the quantities computed throughout the paper, none of our major conclusions are affected.  

The Galactic column density towards NGC4342 is $N_H=1.6 \times 10^{20} \ \rm{cm^{-2}}$ \citep{dickey90}. The $D_{25}$ ellipse of NGC4342 has major and minor axis radii of $37.8\arcsec$ ($4.2$ kpc) and $19.4\arcsec$  ($2.2$ kpc), and the major axis position angle is $166.5\degr$.  Throughout the paper we assume $H_0 = 71  \ \rm{km \ s^{-1} Mpc^{-1}},  \Omega_M=0.3$, and $\Omega_{\Lambda}=0.7$. The errors quoted in the paper are $1\sigma$ uncertainties.

The paper is structured as follows: in Sect. 2 we introduce the X-ray and K-band observations and describe their reduction. In Sect. 3 we study the nature and physical properties of the diffuse X-ray emission, and investigate the dynamics of NGC4342. The population of resolved point sources is examined in Sect. 4.  The results are summarized in Sect. 5.

\section{Data reduction}
\subsection{Chandra}
\label{sec:chandra}
NGC4342 was observed twice by \textit{Chandra} on 2005 February 11 for 38.8 ks (Obs ID: 4687) and on 2011 February 17 for 73.7 ks (Obs ID: 12955) with the ACIS-S detector. The data reduction was performed using standard \textsc{CIAO} software package tools (\textsc{CIAO} version 4.3; \textsc{CALDB} version 4.4.2). The main steps in the  data analyis are those outlined in \citet{bogdan08}. After filtering flare contaminated time intervals, the total remaining exposure time was 78.8 ks. The data were combined by projecting the shorter observation into the coordinate system of Obs ID 12955. Point sources were detected with the CIAO \textsc{wavdetect} tool in the $0.5-8$ keV energy range. For the point source detection the unfiltered merged data was used, the longer exposure time, obtained without omitting periods of high background, allowed a more sensitive detection of point sources.  The spatial scales on which we searched for sources were the  $\sqrt2$-series from 1.0 to 8.0. Additionally, we changed a number of parameters from their default values\footnote{For detailed description of the modified parameters see http://cxc.harvard.edu/ciao/}, in particular we set the value of \textsc{eenergy} to $0.5$,  \textsc{maxiter} to $10$, \textsc{iterstop} to $10^{-5}$, and \textsc{ellsigma} to $4$. As a result we obtained relatively large source cells, which include $\gtrsim98\%$  of the source counts. The resulting source list was used to study the population of bright X-ray sources and to mask out the point sources for further study of the diffuse emission. 

\begin{figure}[t]
    \leavevmode\epsfxsize=8.5cm\epsfbox{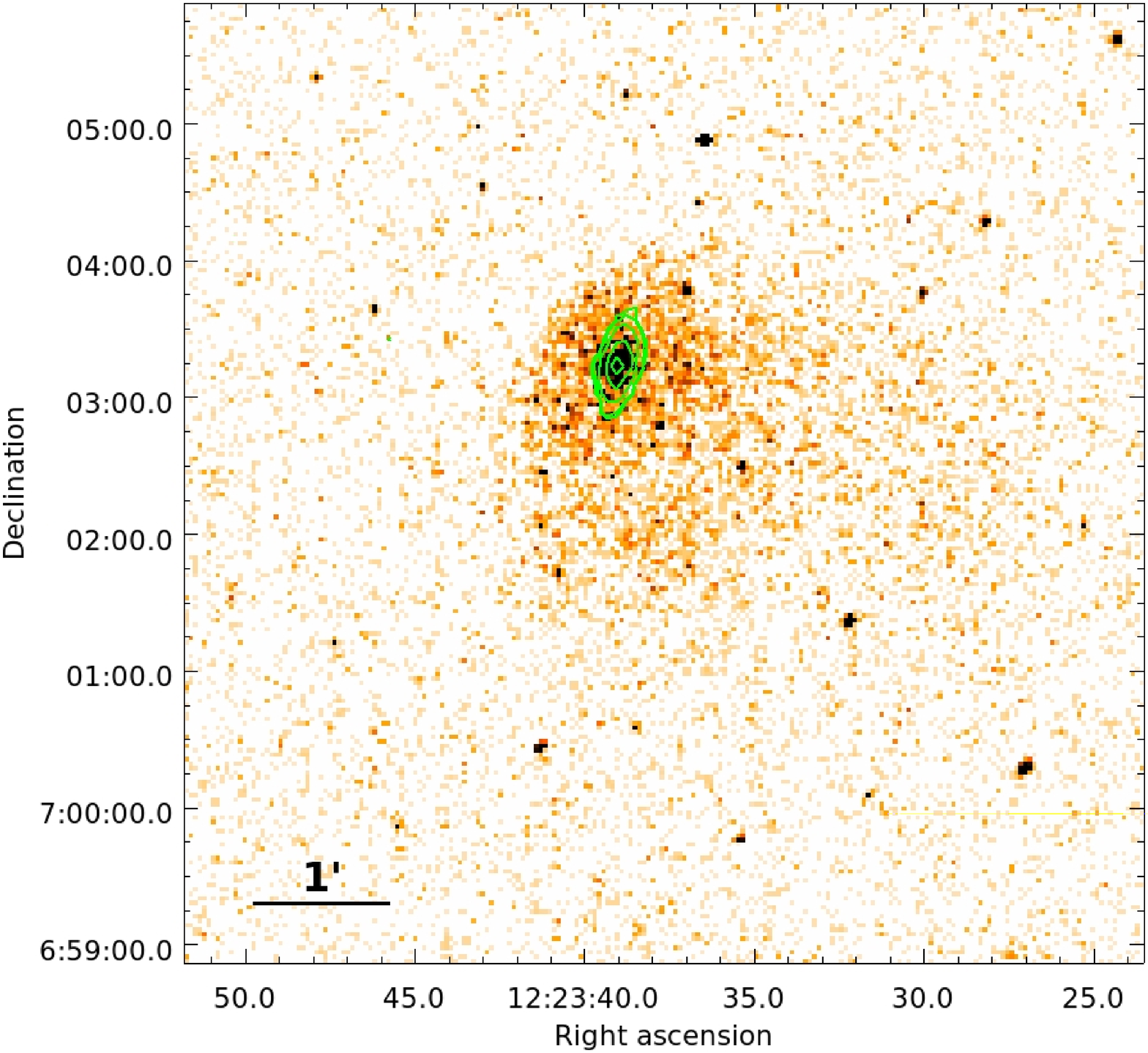}
    \caption{\textit{Chandra} ACIS-S3 image of NGC4342 and its surroundings in the $0.3-1.5$ keV energy range. Contours show isophotes of the galaxy in the K-band light. The X-ray emission associated with NGC4342 has a significantly broader distribution than the stellar light. The diffuse emission shows a sharp surface brightness edge at the northeastern side and a trailing tail at the southwestern side of the galaxy.}
\label{fig:chandra_soft}
\end{figure}

\begin{figure*}
  \begin{center}
    \leavevmode
       \epsfxsize=8.5cm\epsfbox{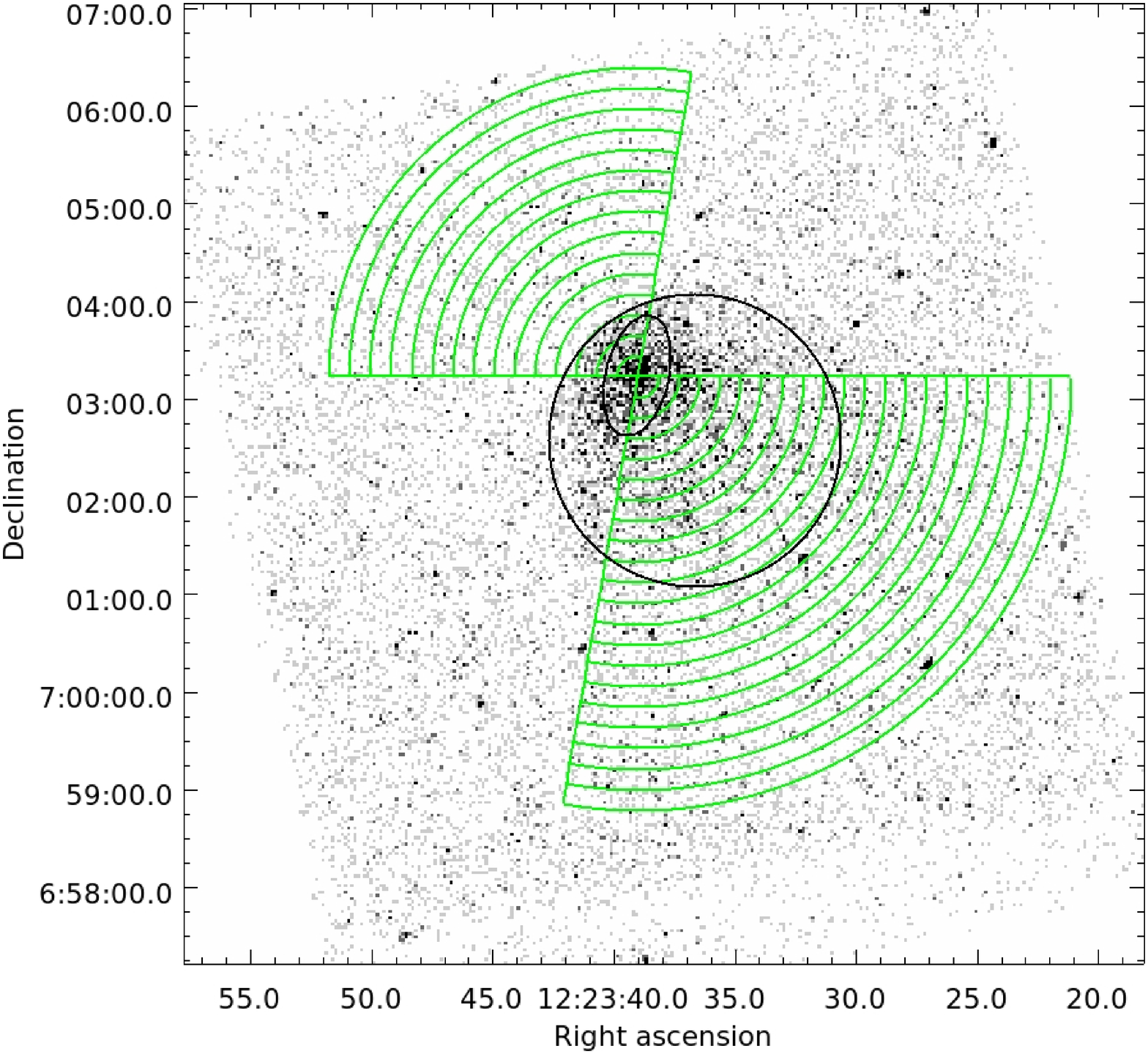}
\hspace{0.75cm} 
      \epsfxsize=8.5cm\epsfbox{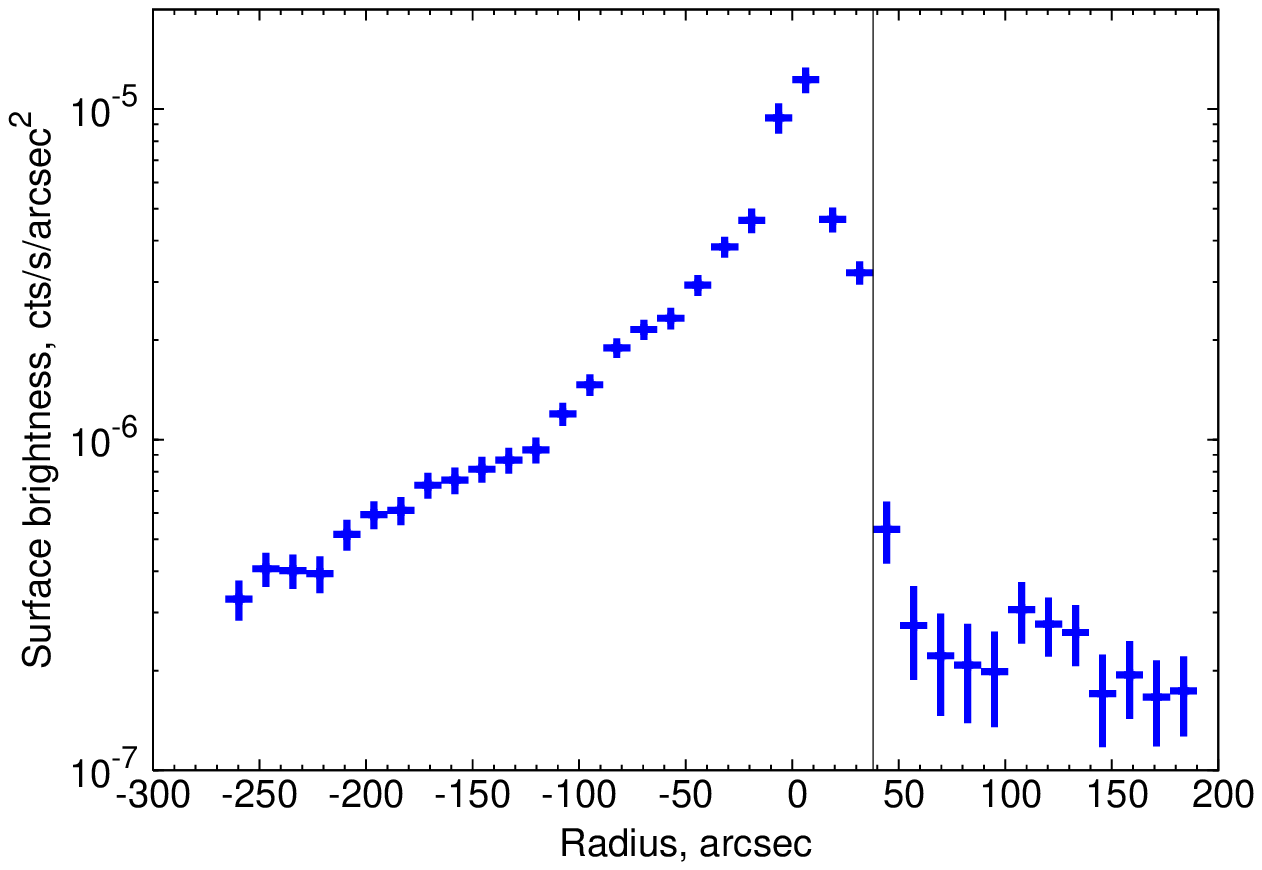}
      \caption{\textit{Left:} $0.5-2$ keV band \textit{Chandra} image of NGC4342. Overplotted circular wedges were used to obtain the surface brightness profile. The position angles of the wedges are $80-180\degr$ towards the northeast and $260-360\degr$ towards the southwest. The small elliptical region shows the $D_{25}$ ellipse. The larger circular region with  $90\arcsec$  radius was used for spectral extraction.  \textit{Right:} Surface brightness profile of the $0.5-2$ keV band diffuse emission. The background is subtracted and the vignetting correction is applied.  At the position of the edge (at $38\arcsec$ central distance, marked by the thin vertical line) the surface brightness drops by a factor of about $6$. The $x$-coordinate increases from southwest to northeast.}
     \label{fig:profile}
  \end{center}
\end{figure*}

To correct for vignetting and to estimate source detection sensitivities, exposure maps were produced using a power law model for the source emission with a slope of  $\Gamma=1.56$, typical for LMXBs \citep{irwin03}. Assuming this spectrum and 7 photons as a detection threshold, the  source detection sensitivity of the combined observation is $\sim$$3\times10^{37} \ \rm{erg \ s^{-1}}$. The X-ray energy spectra of the diffuse emission were extracted using the \textsc{specextract} tool of \textsc{CIAO}.

For the study of the unresolved emission, the use of local background was not possible, since the diffuse emission fills the entire field-of-view (FOV). Therefore we relied on ``blank-sky'' observations\footnote{http://cxc.harvard.edu/contrib/maxim/acisbg/}  to subtract the sky and instrumental background components. To account for variations in the normalization of the instrumental background, we normalized the background level using the $10-12$ keV band count rate ratios. 

The temperature map was derived using the method described by \citet{randall08}. For each temperature map pixel, we extracted a spectrum from a circular region containing a minimum of $400$ net counts in the $0.3-2$ keV energy range. The resulting spectra were fitted with an absorbed APEC model with \textsc{Xspec}, with an abundance fixed at $0.4$ Solar \citep{grevesse98}.

\subsection{Two-Micron All Sky Survey}
Since the near-infrared images of the Two-Micron All Sky Survey (2MASS) Large Galaxy Atlas (LGA) \citep{jarrett03} are excellent stellar mass tracers, we relied on the 2MASS K-band data for this purpose. The K-band images of NGC4342 are not background subtracted, therefore nearby regions off the galaxy were used to estimate the background level. To convert the counts of the K-band image to luminosity, we assumed that the absolute K-band magnitude of the Sun is $ M_{K,\odot} = 3.28 $ mag. The total K-band luminosity of NGC4342  within the $D_{25}$ ellipse is  $L_K = 2.7\times 10^{10} \ \rm{L_{K,\odot}}$.

\section{Diffuse X-ray emission in and around NGC4342}
\subsection{Diffuse X-ray emission in the soft band}
\subsubsection{Hot gas in NGC4342}
\label{sec:diffuse_soft}
The $0.5-2$ keV band \textit{Chandra} image of NGC4342 (Fig. \ref{fig:chandra_soft}) unveils the presence of  diffuse emission associated with the galaxy, which  has a significantly broader distribution than the stellar light. While $\sim$$85\%$ of the K-band  light of NGC4342 is located within an $0.7\times2.2$ kpc elliptical region, the diffuse X-ray emission of NGC4342 extends to $\gtrsim10$ kpc. This  indicates that a large fraction of the unresolved soft-band X-ray emission is not associated with the stellar population but originates from hot ionized gas.

To characterize the physical properties of the diffuse emission, we extract its energy spectrum within the $D_{25}$ ellipse and within a circular region with $90\arcsec$ radius centered at RA: 12h23m36.66s; Dec: +07d02m35.02s (Fig. \ref{fig:profile} left panel). We fit the X-ray spectra with a two component model consisting of an optically-thin thermal plasma emission model (\textsc{APEC} in \textsc{Xspec}) and a power law model. The former component describes the hot X-ray emitting gas, whereas the latter describes the emission from the population of unresolved LMXBs and unresolved faint compact objects (Section \ref{sec:hardband}). The column density was  fixed at the Galactic value and the abundance was set free to vary. The applied model fits the spectra well. Below $\sim$$1.5$ keV energy the spectra are dominated by the thermal component, thereby confirming the gaseous nature of the X-ray emission. The best fit temperature and abundance of the hot gas within the $D_{25}$ ellipse is $kT=0.63\pm0.03$ keV and $0.41^{+0.10}_{-0.07}$ Solar \citep{grevesse98}. Within the larger circular region we obtained similar values, namely $kT=0.64\pm0.02$ keV and $0.47^{+0.24}_{-0.12}$ Solar abundance \citep{grevesse98}. This suggests that the X-ray emitting gas is approximately isothermal within the studied $90\arcsec$ region. From the emission measure ($\int n_e n_h dV$) of the hot gas, the gas mass can be estimated. Within the $D_{25}$ ellipse we obtain $2.5 \times 10^{7} \ \rm{M_{\odot}}$ mass and an average number density of $1.2\times10^{-2} \ \rm{cm^{-3}}$. In the circular region, that includes large fraction of the hot gas associated with NGC4342, we calculate a gas mass of $2.9\times10^{8} \ \rm{M_{\odot}}$ and an average number density of $2.8\times10^{-3} \ \rm{cm^{-3}}$.

From the best-fit spectra we find the total $0.5-2$ keV band X-ray luminosity of the diffuse emission. Within the $D_{25}$ ellipse we obtain $2.7\times10^{39} \ \rm{erg \ s^{-1}}$. Given the total K-band luminosity of NGC4342, $L_K = 2.7\times 10^{10} \ \rm{L_{K,\odot}}$, the X-ray-to-K-band luminosity ratio of the galaxy is $1.0\times10^{29} \ \rm{erg \ s^{-1} \ L_{K,\odot}^{-1}}$. Within the circular region with $90\arcsec$ radius, we obtain $1.0\times10^{40} \ \rm{erg \ s^{-1}}$ luminosity in the $0.5-2$ keV band, which results in the $L_X/L_K$ ratio of $3.7\times10^{29} \ \rm{erg \ s^{-1} \ L_{K,\odot}^{-1}}$. The luminous nature of the diffuse emission and the high $L_X/L_K$ ratios also indicate that the diffuse emission is dominated by  hot gas and only a small fraction of this component arises from unresolved compact objects. In particular, the $L_X/L_K$ ratio of unresolved faint compact objects (ABs and CVs) is in the range of $(3-8)\times10^{27} \ \rm{erg \ s^{-1} \ L_{K,\odot}^{-1}}$ in the $0.5-2$ keV band \citep{bogdan11}. On the other hand, unresolved LMXBs ($L_X < 3 \times 10^{37} \ \rm{erg \ s^{-1}}$) are expected to contribute with an $L_X/L_K$ ratio of $\sim$$1.0\times10^{28} \ \rm{erg \ s^{-1} \ L_{K,\odot}^{-1}}$ in the $0.5-2$ keV band. To obtain this value we  assume the average LMXB luminosity function \citep{gilfanov04} and the average LMXB power law spectrum \citep{irwin03}. Thus, unresolved compact objects can account only for $\lesssim5\%$ of the total X-ray emission in the large circular region  and for $\lesssim20\%$ within the $D_{25}$ ellipse.

The observed $L_X/L_K$ ratios in NGC4342  are atypical of low-stellar mass galaxies ($\lesssim$$10^{11} \ \rm{M_{\odot}}$), but are observed in ellipticals that have $\sim$$10-30$ times more stellar mass \citep{bogdan11}.  \textit{Although this result may be initially unexpected, it actually  demonstrates that the amount of X-ray emitting gas in galaxies depends on their total gravitating mass rather than solely on their stellar mass.} Indeed, in \citet{bogdan12} we pointed out that NGC4342 has a massive dark matter halo, with a  total gravitating mass in the range of $(1.4-2.3)\times 10^{11} \ \rm{M_{\odot}}$  within $10$ kpc, which is typical of galaxies with $\sim$$10-30$ times more stellar mass \citep{buote02,humphrey06}. We therefore conclude that neither the presence of nor the observed physical properties of the bright X-ray halo are unusual, given the massive dark matter halo surrounding NGC4342.  Moreover, the case of NGC4342 points out that part of the scatter observed in the $L_X-L_K$ (or $L_X-L_B$) relation \citep[e.g.][]{osullivan01,bogdan11} of early-type galaxies may be due to the fact that galaxies with similar stellar mass can have markedly different dark matter halo masses.

\subsubsection{The X-ray surface brightness edge}
\label{sec:edge}
The X-ray light distribution of NGC4342 shows strong hydrodynamic disturbances with a very distinct surface brightness edge at the northeastern side and a following tail at the southwestern side of the galaxy (Fig. \ref{fig:chandra_soft}). To investigate the head-tail structure of the hot gas, we construct a surface brightness profile and map the temperature distribution of the diffuse emission. 

We depict the $0.5-2$ keV surface brightness distribution in the right panel of Fig. \ref{fig:profile}. The circular wedges  used to extract the profile are shown in the left panel of Fig. \ref{fig:profile}. The profiles are corrected for vignetting and the background is subtracted. As indicated by the soft band \textit{Chandra} image (Fig. \ref{fig:chandra_soft})  the surface brightness abruptly drops at the northeastern side of the galaxy at about $38\arcsec$ from the galaxy center.  The decrease in surface brightness is approximately a factor of six. Beyond the edge, the  surface brightness is approximately constant, indicating the presence of large-scale X-ray emission outside NGC4342. In the opposite, southwestern direction, the surface brightness decreases continuously, but even at $\sim$$250\arcsec$  is  higher than that measured on the northeastern side of NGC4342. To sum up, the surface brightness distribution confirms the head-tail distribution of the X-ray emitting gas, and indicates that the galaxy is surrounded by large-scale diffuse emission. 

\begin{figure}[t]
    \leavevmode\epsfxsize=8.5cm\epsfbox{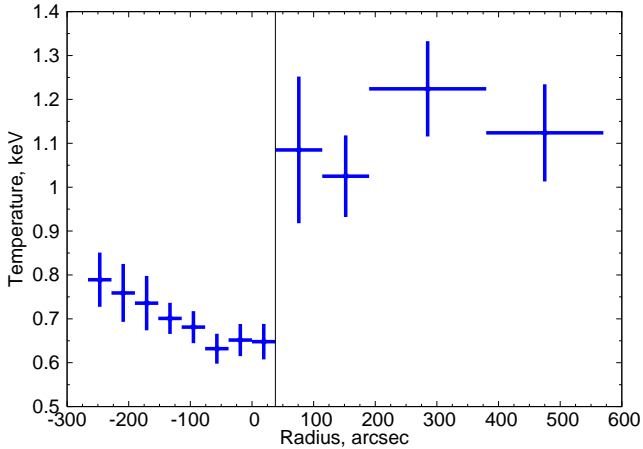}
    \caption{Temperature profile of the diffuse emission in and around NGC4342. The spectra were extracted with similar wedges used to obtain Fig. \ref{fig:profile}. The position of the surface brightness edge is marked with the thin vertical line. The last two points were obtained using the ACIS-S2 detector.}
\label{fig:temperature}
\end{figure}

\begin{figure}[t]
    \leavevmode\epsfxsize=8.5cm\epsfbox{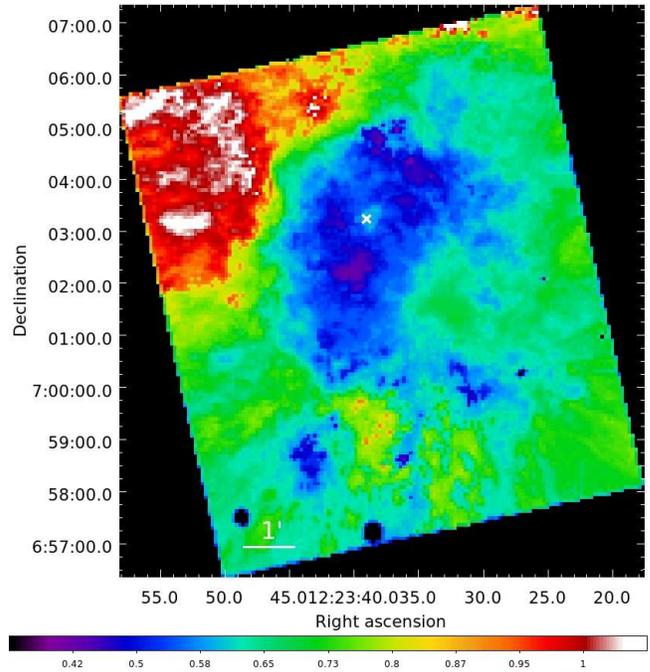}
    \caption{Temperature map of NGC4342 and its environment derived from ACIS-S3 data. The color-bar gives the
temperature in keV. The cross marks the center of NGC4342. The temperature within the central $\sim$$90\arcsec$ region is fairly isothermal but shows a sharp increase at the approximate position of the surface brightness edge.}
\label{fig:tmap}
\end{figure}

To map the temperature distribution of the hot X-ray emitting gas, we used two approaches. First, we studied the temperature distribution along the surface brightness edge and tail using wedges with the same position angles as for the profiles. For each region we extracted an X-ray  spectrum and fit it using an optically-thin thermal plasma emission model (\textsc{APEC} in \textsc{Xspec}) with column density fixed at the Galactic level and abundance fixed at $0.4$ Solar \citep{grevesse98}. The temperature distribution, shown in  Fig. \ref{fig:temperature}, demonstrates that the X-ray emitting gas associated with NGC4342 is fairly isothermal with a temperature of $kT\sim0.6$ keV. Outside the galaxy, at the position of the surface brightness edge the gas temperature sharply increases to  $\sim$$1$ keV.  At the southwestern side of the galaxy beyond $90\arcsec$ ($10$ kpc in projection) the temperature gradually rises from $\sim$$0.6$ keV to $\sim$$0.8$ keV. The low temperature of the X-ray gas in the tail suggests that it mostly originates from galaxy gas rather than from the hotter group gas. As a second approach, we produced a temperature map  of NGC4342 and its neighborhood using the ACIS-S3 detector (Fig. \ref{fig:tmap}). The temperature map, in good agreement with the temperature profile, demonstrates that the hot gas is approximately isothermal within $\sim$$90\arcsec$  -- its temperature is $\sim$$0.6$ keV. Moreover, Fig. \ref{fig:tmap}  illustrates the presence of a sharp increase in the temperature at the approximate location of the surface brightness edge. We note that the offset between the actual position of the edge and the temperature increase in  Fig. \ref{fig:tmap} is the consequence of the method used to construct the temperature map.  

From the observed surface brightness distribution and the temperature difference inside and outside the edge, we conclude that the detected sharp surface brightness edge is a contact discontinuity or cold front. Thus, the observed head-tail distribution of the hot gas is produced by ram pressure as NGC4342 moves through the external gas \citep{vikhlinin01,markevitch07}.

\subsubsection{Large-scale diffuse X-ray emission around NGC4342}
\label{sec:groupgas}
Diffuse  X-ray emission is not only associated with NGC4342, but also extends to large scales  around the galaxy.  To study the external diffuse emission, we extracted its X-ray energy spectrum using three off-axis \textit{Chandra} detectors, namely ACIS-S2, ACIS-I2, and ACIS-I3. For each detector the extraction region was a $400\arcsec\times400\arcsec$ rectangle centered on the center of the CCD. The resulting spectra were fit with an  optically-thin thermal plasma model (\textsc{APEC} model in \textsc{Xspec}) with abundance  fixed at $0.4$ Solar \citep{grevesse98}. The spectral fits are statistically acceptable, which suggests that the  emission originates from hot group/cluster gas.  The  best-fit temperatures are in the range of $kT=1.03-1.20$ keV and agree within statistical uncertainties. Given the best-fit parameters, we also compute the average surface brightness of the external emission and found it to be  fairly uniform. On each off-axis CCD the surface brightness level is between $ (2.4-3.0)\times10^{-15} \ \rm{erg \ s^{-1} \ cm^{-2} \ arcmin^{-2}} $. 

The fairly uniform properties of the external gas out to $\sim$$15\arcmin$ ($\sim$$100$ kpc in projection) distance from NGC4342 suggest that this emission has a large extent compared to the galaxy. Given the proximity of NGC4365 to NGC4342, which  is presumably the center of the $W'$ cloud, it is plausible that a large fraction of the observed external diffuse emission  is associated with the $W'$ cloud. Since NGC4342 lies  $130$ kpc in projection to the southwest from NGC4365, the existence of the group gas around NGC4342 places a lower limit on the extent of the $W'$ cloud. We note that according to the study of \citet{mei07}  the extent of the $W'$ cloud may be much larger than this lower limit, since two galaxies (NGC4434 and NGC4318), presumably belonging the group, lie at a projected distance of $\sim$$ 1\degr$ ($\sim$$400$ kpc in projection) from NGC4365. 

However, the presence of the underlying Virgo cluster emission introduces a major caveat in exploring  the properties  of the group gas. At the position of NGC4342, the temperature of the Virgo cluster gas may be similar or somewhat higher than  $kT\sim$$1.1$ keV, furthermore its abundance may be also close to $0.4$ Solar. Therefore, the hot X-ray emitting gas, originating from  the $W'$ cloud is virtually indistinguishable from the Virgo cluster emission. This effect has two major consequences: First, the  best-fit parameters do not necessarily characterize the X-ray emitting gas of the $W'$ cloud, as it could be blended with the underlying Virgo emission. For this reason the physical conditions of the group gas, such as temperature, density, pressure,  cannot be measured precisely. Second, the surface brightness profile of the emission also cannot be characterized accurately, since beyond the virial radius of the Virgo cluster ($r_{\rm{200}}\sim3.9\degr$) the distribution of the cluster gas is largely unknown. 

Acknowledging these uncertainties, we provide  an admittedly crude estimate on the total gravitating mass of the $W'$ cloud within $130$ kpc. We assume that the group gas is isothermal and has a temperature of $kT=1.1$ keV, furthermore we assume that the X-ray gas has a spherically symmetric distribution.  Using these assumptions we estimate that the total gravitating mass within $130$ kpc is $\sim$$5\times10^{12} \ \rm{M_{\odot}}$. As discussed above, the $W'$ cloud is most likely significantly larger than $130$ kpc, therefore this mass estimate should be considered as a lower limit. To explore the spatial distribution and better understand  the physical properties  of the  X-ray gas associated with the $W'$ cloud further X-ray observations would be highly beneficial.

\subsubsection{The velocity of NGC4342}
The detection of a cold front  and the presence of large-scale diffuse emission implies that  NGC4342 moves with high velocity through an ambient medium towards the northeast, i.e. towards NGC4365. The velocity of NGC4342 can be estimated using the ratio of the thermal pressures at the stagnation point ($p_0$) and in the free stream ($p_1$) \citep{vikhlinin01}. However, several uncertainties are involved in this calculation. First, the path length of the external gas is rather uncertain. According to \citet{mei07} the $W'$ cloud may extend out to $\sim$$400$ kpc (Sect. \ref{sec:groupgas}), which is typical for a poor  galaxy group.  Second, the thermal pressures are measured in relatively large regions hence we assume that no pressure gradients are present. While this may be a good assumption outside the edge, it somewhat overestimates the pressure inside the edge. Therefore the $p_0/p_1$ ratio hence the velocity of NGC4342 will be slightly overestimated. Finally, due to the underlying Virgo emission (for discussion see Sect. \ref{sec:groupgas}) the properties of the grup gas of the $W'$ cloud cannot be determined with certainty. Therefore, if a significant fraction of the external gas originates from the Virgo cluster, the estimated density and hence the pressure of the group gas becomes lower, resulting in a larger $p_0/p_1$ ratio and a larger velocity for NGC4342. 

\begin{figure}[t]
    \leavevmode\epsfxsize=8.5cm\epsfbox{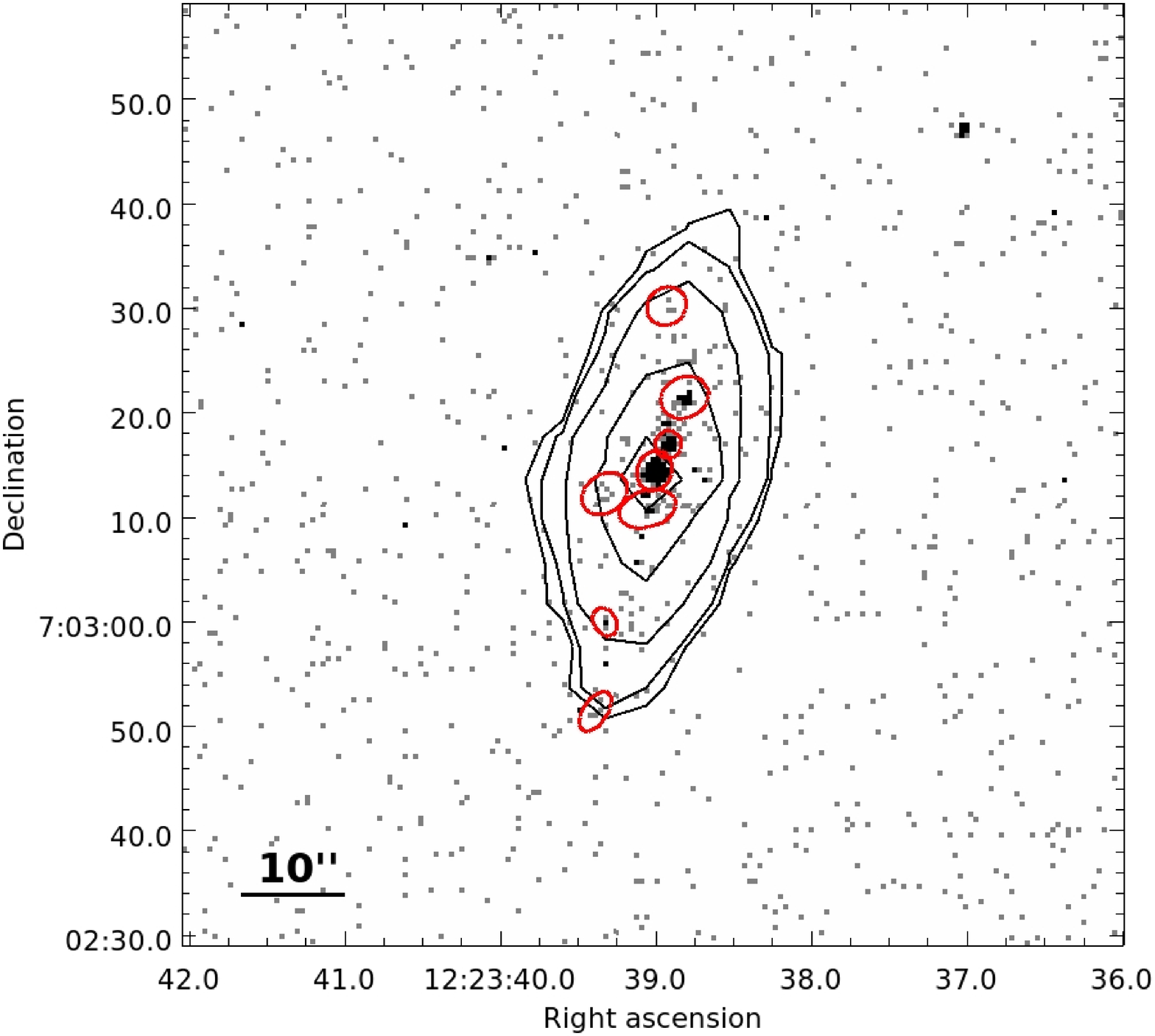}
    \caption{\textit{Chandra} image of NGC4342 in the $1.5-7$ keV energy range. Contours show isophotes of the galaxy in the K-band light. Overplotted elliptical regions show the position and extent of the detected sources. The  location and physical properties of the sources are listed in Table \ref{tab:sources}.}
\label{fig:chandra_hard}
\end{figure}

To compute the pressure ratios, we extract spectra of two circular wedges centered on  NGC4342 with radii of $0-38\arcsec$ and $38-114\arcsec$ and position angle of $80-180\degr$. Note that the surface brightness edge is located at $38\arcsec$ from the galaxy center. For the region inside the edge, we assume spherical symmetry, whereas outside the edge we assume a path length of $400$ kpc.  The spectra were fit with thermal plasma models as described in Section \ref{sec:edge}. The best-fit temperatures inside and outside the edge are $kT_0=0.65$ keV and $kT_1=1.08$ keV. From the emission measure, we compute  the number densities inside and outside the edge and obtain  $1.3\times10^{-2} \ \rm{cm^{-3}}$ and $4.0\times10^{-4} \ \rm{cm^{-3}}$, respectively. The average pressure can be computed as $p=1.9n_ekT$, hence we obtain $p_0=2.6\times10^{-11} \ \rm{erg \ cm^{-3}}$ and $p_1= 1.4\times10^{-12} \ \rm{erg \ cm^{-3}}$, implying a pressure ratio $p_0/p_1=18.6\pm4.8$. This pressure ratio corresponds to a  Mach-number of $M=2.6\pm0.2$  \citet{vikhlinin01}. We stress that the given uncertainties correspond to the statistical uncertainties in the determination of the pressures, and do not represent the above discussed systematic uncertainties. The sound speed in a $kT=1$ keV plasma is $c_s=\sqrt{(\gamma kT)/\mu m_H} = 505 \ \rm{km \ s^{-1}}$ assuming $\gamma=5/3$ and $\mu=0.62$.  Thus, NGC4342 is falling towards NGC4365 with a velocity of $1300\pm260 \ \rm{km \ s^{-1}}$.

Based on the radial velocities of NGC4342 and NGC4365, and on the supersonic ($\sim$$1300 \ \rm{km \ s^{-1}}$) infall velocity of the former galaxy, we determine the orientation of the motion of NGC4342. The radial velocities of NGC4342 and NGC4365 are $751 \ \rm{km \ s^{-1}}$ and $1243 \ \rm{km \ s^{-1}}$, hence the radial velocity difference is $-492 \ \rm{km \ s^{-1}}$. Assuming that this value represents the relative radial velocity between NGC4342 and the ambient group gas, the plane of the sky velocity of NGC4342 is $\sim$$1200 \ \rm{km \ s^{-1}}$. We thus deduce that the direction of motion of NGC4342 is $\sim$$22\degr$ relative to the plane of the sky. 

\begin{table*}
  \caption{Properties of point sources detected within the $D_{25}$ ellipse of NGC4342.}
  \label{tab:sources}
\renewcommand{\arraystretch}{1.3} 
 \begin{center}
    \leavevmode
    \begin{tabular}{cccccccc} \hline \hline              
  ID &   CXOU Name    &    R.A. & Dec. & Central distance & Source counts$^{\dagger}$ & Background counts$^{\dagger}$ & $L_{0.5-8 \rm{keV}}$$^{\ddagger}$      \\ 
    &  &  (J2000)     &  (J2000) & ($''$)   &            &      & ($\rm{erg \ s^{-1}}$)                               \\ \hline
1   & J122339.0+070314 & 12:23:39.005 & +07:03:14.38 & 0.3 &  442 &   7.5 & $2.6\times 10^{39}$ \\
2   & J122338.9+070316 & 12:23:38.916 & +07:03:16.93 & 3.2   & 177&   3.9 & $1.1\times 10^{39}$ \\
3   & J122339.0+070310 & 12:23:39.049 & +07:03:10.77 & 3.4   &  50 & 11.3 & $2.4\times 10^{38}$ \\
4   & J122339.3+070312 & 12:23:39.324 & +07:03:12.21 & 4.9   &  34 &   9.8 & $1.5\times 10^{38}$\\
5   & J122338.8+070321 & 12:23:38.810 & +07:03:21.42 & 7.9   &  60 &10.6 & $3.1\times 10^{38}$ \\
6   & J122339.3+070259 & 12:23:39.323 & +07:02:59.95 & 14.9 &  12 & 2.1  & $6.1\times 10^{37}$\\
7   & J122338.9+070330 & 12:23:38.927 & +07:03:30.15 & 16.0 &  12 & 7.1  & $2.8\times 10^{37}$ \\ 
8   & J122339.3+070251 & 12:23:39.384 & +07:02:51.38 & 23.4 & 11  & 5.0   & $3.7\times 10^{37}$\\ \hline
\multicolumn{8}{l}{} \\
\multicolumn{8}{l}{$^{\dagger}$ The number of source and background counts refer to the $0.5-8$ keV energy range.}\\
\multicolumn{8}{l}{$^{\ddagger}$ Calculated assuming the average LMXB spectrum \citep{irwin03} except for source (1) where its best-fit spectrum was used.}
    \end{tabular}
  \end{center}
\end{table*}

\subsection{Diffuse emission in the hard X-ray band}
\label{sec:hardband}
Within the optical extent of NGC4342, we also observe diffuse X-ray emission in the hard energy band. To compute the total luminosity of the diffuse emission in the $2-10$ keV energy range, we rely on the spectrum extracted for the $D_{25}$ ellipse. The full band spectrum is described with the a two component model consisting of a thermal model and a power law model (Sect. \ref{sec:diffuse_soft}). Since the thermal component does not play a notable role above $\sim$$1.5$ keV energy, the hard tail of the spectrum is described by the power law model. The best-fit slope of the power law is $\Gamma=1.83^{+0.34}_{-0.24}$, which is in good agreement with those found for other early-type galaxies \citep{bogdan11}. Using the best-fit parameters we compute that the $2-10$ keV band luminosity of the diffuse emission is $7.5\times10^{38} \ \rm{erg \ s^{-1}}$. Given the K-band luminosity of NGC4342 the $L_X/L_K$ ratio  within $D_{25}$ ellipse is $2.8\times10^{28} \ \rm{erg \ s^{-1} \ L_{K,\odot}^{-1}}$. 

The diffuse X-ray light in the hard band arises presumably from the combined emission of unresolved LMXBs and faint compact objects (ABs and CVs). We estimate the contribution of unresolved LMXBs using their average luminosity function \citep{gilfanov04} following the method described in Sect. \ref{sec:diffuse_soft}. Below the source detection threshold, we expect $2.4\times10^{28} \ \rm{erg \ s^{-1} \ L_{K,\odot}^{-1}}$. Faint unresolved compact objects contribute with $(3-4)\times10^{27} \ \rm{erg \ s^{-1} \ L_{K,\odot}^{-1}}$ in the $2-10$ keV energy band \citep{bogdan11}. Thus, the combination of these two components is in good agreement with the observed $L_X/L_K$ ratios. Hence the bulk of the $2-10$ keV band diffuse  emission originates from  the population of unresolved LMXBs and unresolved faint compact objects, mainly ABs and CVs.

\section{Resolved X-ray point sources}
\subsection{Resolved sources within the $D_{25}$ ellipse}
Within the $D_{25}$ ellipse of NGC4342, we detect 8 bright point sources. In Table \ref{tab:sources} we list the position, the number of total and background counts within the source cells, and the luminosity of the sources in the $0.5-8$ keV energy band.  To estimate the background level at the source locations we used surrounding annuli or nearby circular apertures if another source was lying too close. 

The most luminous source is coincident with the center of NGC4342. To study its nature we extracted its X-ray energy spectrum and fit it with a  power law model with the column density fixed at the Galactic value. The resulting fit is acceptable, the best-fit slope of the power law model is $\Gamma=1.42\pm0.07$. According to the best-fit model, the luminosity of this source is $2.6\times 10^{39} \ \rm{erg \ s^{-1}}$. Based on the position of the nuclear source, we conclude that it is associated with the supermassive black hole of NGC4342. 

To study the $7$ non-central point sources, we extracted their composite X-ray energy spectrum and fit it with a power law model assuming Galactic column density. The resulting acceptable fit yields a power law slope of $\Gamma=1.60\pm0.08$. Although not conclusive, the best-fit model is consistent with the typical LMXB spectrum \citep{irwin03}. To estimate the number of cosmic X-ray background (CXB) sources within the $D_{25}$ ellipse, we rely on the $\log N - \log S$ distribution of \citet{georgakakis08}. To convert the $0.5-10$ keV band $\log N - \log S$ distribution of CXB sources to the $0.5-8$ keV energy range, we assumed a power law model with a slope of $\Gamma=1.4$. Additonally, we also corrected for incompleteness effects using the procedure described in \citet{zhang12}. Within the  $D_{25}$ ellipse, the CXB source density is $\approx$$0.56 \ \rm{arcmin^{-2}}$, hence we expect $\approx$$0.36$ CXB sources. We thus conclude that majority ($\sim$$6-7$) of the non-central resolved sources are LMXBs.

\begin{figure*}
  \begin{center}
    \leavevmode
       \epsfxsize=8.5cm\epsfbox{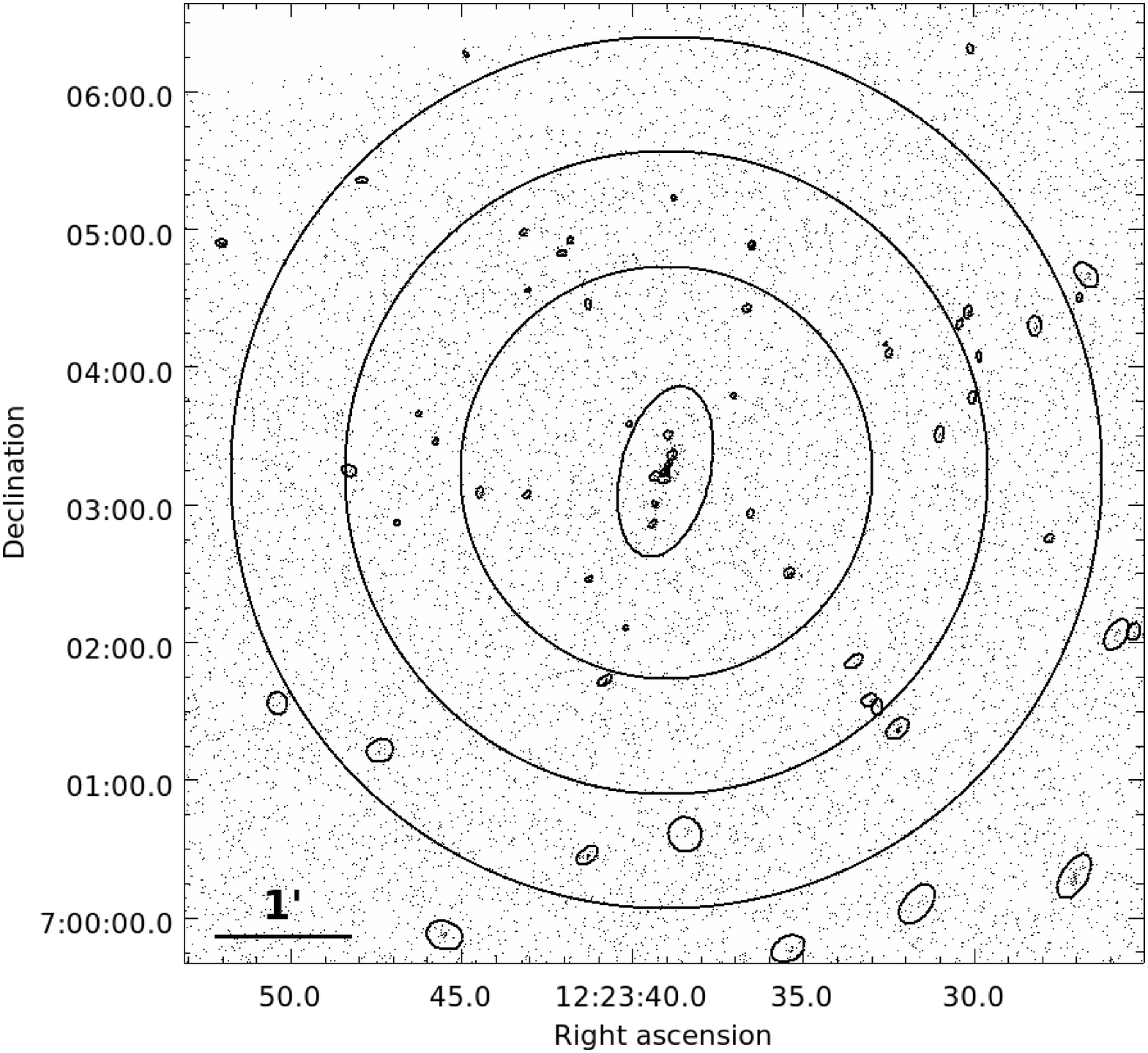}
\hspace{0.75cm} 
      \epsfxsize=8.5cm\epsfbox{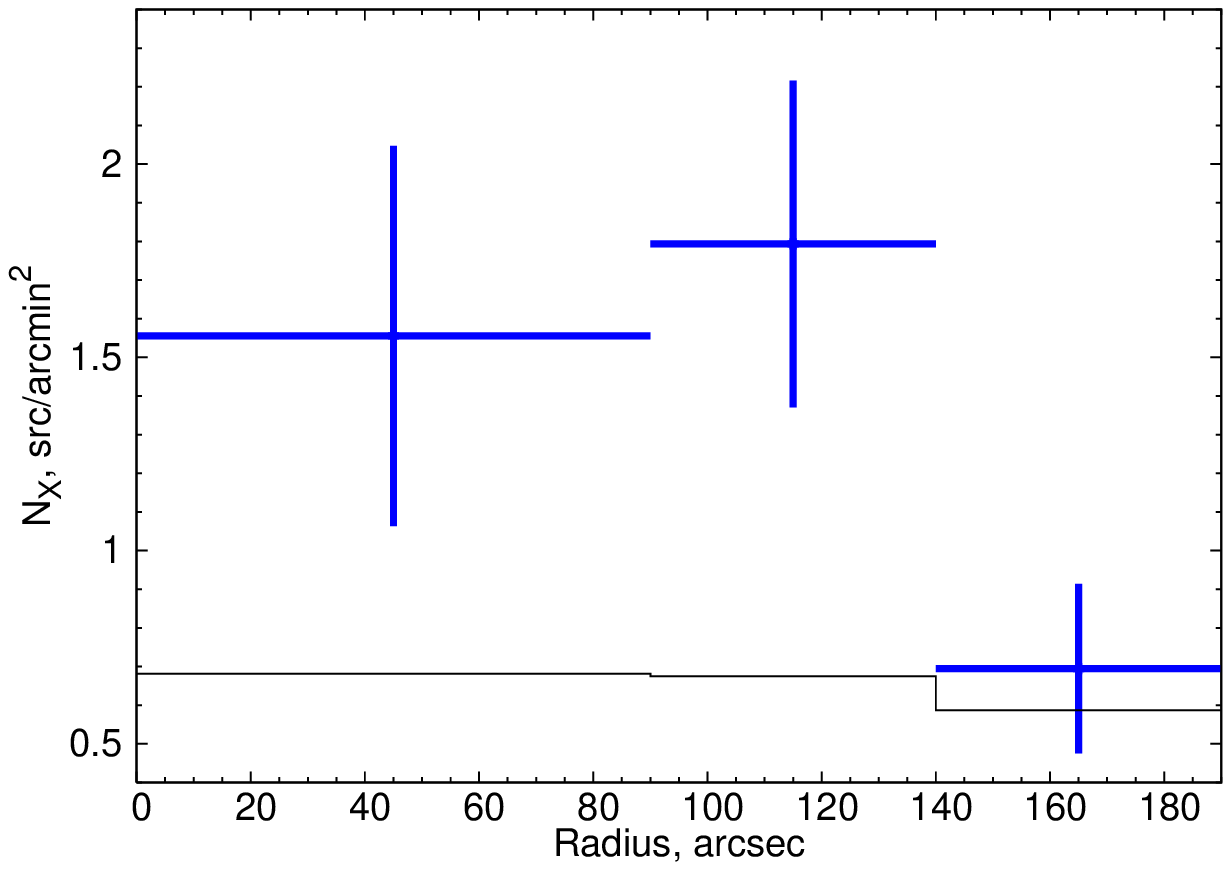}
      \caption{\textit{Left:} Hard band ($1.5-7.5$ keV) image of NGC4342 and its surroundings. Bright resolved sources are marked with elliptical regions. The number of sources and source densities were computed within the overplotted circular annuli. Point sources within the $D_{25}$ ellipse (shown within the inner circular region) were excluded.  \textit{Right:} Density of resolved sources in the three annuli around NGC4342. The solid line shows the expected number of CXB sources within each annulus.}
     \label{fig:source_excess}
  \end{center}
\end{figure*}

\subsection{Resolved X-ray sources outside the $D_{25}$ ellipse}
Outside the $D_{25}$ ellipse we resolved a large number of point sources. To study the spatial distribution of the detected sources, we divided the surroundings of NGC4342 into three circular annuli centered on the center of the galaxy. The outer radii of the annuli were $90\arcsec$, $140\arcsec$, and $190\arcsec$, respectively (Fig. \ref{fig:source_excess} left panel). Note, that sources within the $D_{25}$ ellipse were excluded. In each annulus we compute the source density and compare it with the expected CXB source density, which was computed based on the $\log N - \log S$ distribution of \citet{georgakakis08}. As before we assumed a power law model with slope of  $\Gamma=1.4$ to convert the $0.5-10$ keV band $\log N - \log S$ distribution to the $0.5-8$ keV energy range, and we applied incompleteness correction following the method of \citet{zhang12}. 

The right panel of Fig. \ref{fig:source_excess} shows that within $140\arcsec$ ($\sim$$15.5$ kpc in projection) of the center of NGC4342 the observed source density significantly exceeds the expected CXB source density. In particular, within $140\arcsec$ we resolve $28$ sources, whereas the predicted number of background sources is $\approx$$11.1$. The $\approx$$16.9$ excess sources correspond to a $\sim$$5\sigma$ excess. Beyond a radius of $140\arcsec$ the  numbers of detected and expected sources are in good agreement. Since the CXB level fluctuates only at the $10-30\%$ level, it is very likely that the excess sources are associated with NGC4342. 

In \citet{bogdan12} we demonstrated that NGC4342 resides in a massive dark matter halo, which extends out to at least $10 $ kpc and possibly much farther. Thus, if the excess X-ray sources are associated with NGC4342, they are located in its dark matter halo, outside the optical extent of the galaxy. In this interpretation, it is feasible that these sources are LMXBs located in globular clusters (GC-LMXBs). Studies of GC-LMXBs in nearby galaxies pointed out that only a few per cent of GCs host LMXBs brighter than $10^{37} \ \rm{erg \ s^{-1}}$ \citep[e.g.][]{angelini01,kundu02,zhang11}. Assuming that $1-2\%$ of blue (metal-poor) GCs host LMXBs brighter than $\sim$$3\times10^{37} \ \rm{erg \ s^{-1}}$, furthermore attributing the $\sim$$17$ excess sources to GC-LMXBs, we estimate that NGC4342 may host $N_{\rm{GC}}\sim850-1700$ GCs. 

The possibly large GC population is further supported by the rather tight correlation between the mass of the central supermassive black hole ($M_{\bullet}$) and the number of GCs in early-type galaxies \citep{burkert10}. Given the best-fit $M_{\bullet}-N_{\rm{GC}}$ relation \citep{burkert10} and the black hole mass \citep{cretton99} of NGC4342 ($4.6\times 10^8 \ \rm{M_{\odot}}$ -- converted to a distance of $23 $ Mpc) the predicted number of GCs is $N_{\rm{GC}}\sim1500$, in good agreement with that predicted from the number of excess X-ray sources. 

The presence of a notable GC population around NGC4342 can be  confirmed by deep optical observations. To probe the GC systems around NGC4342, we rely on the publicly available Canada France Hawaii Telescope / MegaCam (CFHT/MegaCam) imaging and data products (Blom et al. in preparation). The region of NGC4342 is covered with the $u'$ filter image stack (total 4240 s of exposure and $1.3\arcsec$ spatial resolution) and the $g'$ and $i'$ filter stacks (total 3170/2055 s of exposure and $0.93\arcsec/0.60\arcsec$ spatial resolution), which allow the identification of GCs using their locus in $u'-g'$ versus $g'-i'$ colour space. We define this locus with objects also in the NGC4365 GC catalogue derived from Subaru / SuprimeCam (SCam) imaging \citep{blom12}. The spatial distribution of detected GC candidates recover the concentration of GCs around NGC4365 and also demonstrates an overdensity of GCs centred on NGC4342. To give a crude estimate on the number of GC candidates associated with NGC4342, we compare the number of observed GC candidates observed by the CFHT/MegaCam imaging around NGC4365 and NGC4342, and use the lower limit on the number of GCs in NGC4365 \citep{blom12}. We conclude that NGC4342 may host $N_{\rm{GC}}=1200 \pm 500$ GCs. Thus, optical observations are in good agreement with that deduced from the number of resolved X-ray sources. The analysis of the GC population of NGC4342 will be discussed in full particulars in Blom et al. (in preparation). The connection between the GC population and the resolved LMXBs is beyond the scope of the present work and will be studied in a forthcoming paper. 

We emphasize that the estimated number of GCs associated with NGC4342 exceeds by about an order of magnitude the expected value, based on the optical luminosity of NGC4342. To further illustrate this point, we compute the GC specific frequency using \citet{harris81} and the absolute V-band magnitude of NGC4342 ($M_V=-19.45$ mag). We obtain a GC specific frequency of $S_{\rm{N}} = 19.9\pm8.3$, which is one of the largest values reported for full-size galaxies \citep[e.g.][]{peng08}. 

The color distribution of GCs is bimodal: blue metal-poor and red metal-rich sub-populations exist \citep[see][and references therein]{brodie06}. Whereas the distribution of the metal-rich population follows the stellar light, the metal-poor population has a broader distribution  \citep{bassino06} and traces dark matter halos. Thus, the GC population of NGC4342 is likely to be dominated by metal-poor GCs. If optical studies confirm the presence of a notable metal-poor GC population, their formation mechanism  can be constrained. Specifically, the presence of a large metal-poor GC population in NGC4342 -- a galaxy with very low stellar mass but a massive dark matter halo -- would promote an evolutionary scenario, in which  metal-poor GCs are formed prior to the main stellar body of the galaxy. This would be consistent with the formation scenario described by \citet{beasley02}, who -- based on a semianalytic galaxy formation model \citep{cole00} -- suggested that metal-poor GCs are formed at high redshift ($z > 5$) in gas disks in low-mass dark-matter halos before the major burst of star-formation.

\section{Conclusions}
We studied the diffuse X-ray emission and the population of resolved sources in NGC4342 using \textit{Chandra} X-ray observations. Our results can be summarized as follows: \\

1. NGC4342 hosts a bright X-ray halo originating from hot gas with a temperature of $kT\sim0.6$ keV. The distribution of the hot gas is significantly broader than that of the stellar light. Within the $D_{25}$ ellipse, the $0.5-2$ keV band luminosity of the diffuse emission is $2.7\times10^{39} \ \rm{erg \ s^{-1}}$, which yields an $L_X/L_K$ ratio of $1.0\times10^{29} \ \rm{erg \ s^{-1} \ L_{K,\odot}^{-1}}$. 

2. We detect a sharp surface brightness edge at the northeastern side of NGC4342, which we identify as a contact discontinuity or cold front. From the thermal pressure ratios between inside and outside the edge, we estimate that NGC4342 is moving  supersonically ($M\sim2.6$ or $\sim$$1300 \ \rm{km \ s^{-1}}$) towards the northeast, i.e. towards NGC4365. 

3. Within the $D_{25}$ ellipse we resolve 8 point sources. The brightest source has a luminosity of $2.6\times 10^{39} \ \rm{erg \ s^{-1}}$ and is located at the center of NGC4342. Hence we conclude that this source is associated with the supermassive black hole of NGC4342. 

4. Outside the $D_{25}$ ellipse we detect 2$8$ bright X-ray sources, $\sim$$17$ more than expected from the cosmic X-ray background. The excess sources may be LMXBs located in metal-poor GCs, which systems are associated with the massive dark matter halo of NGC4342. Since only a few per cent of GCs host LMXBs brighter than $10^{37} \ \rm{erg \ s^{-1}}$, we estimate that NGC4342 could host  $\sim$$850-1700$ GCs. Optical observations also suggest that NGC4342 may harbor $1200\pm500$ GCs. \\

\begin{small}
\noindent
\textit{Acknowledgements.}
This research has made use of \textit{Chandra}  data provided by the Chandra X-ray Center. The publication makes use of software provided by the Chandra X-ray Center (CXC) in the application package CIAO. This publication makes use of data products from the Two Micron All Sky Survey, which is a joint project of the University of Massachusetts and the Infrared Processing and Analysis Center/California Institute of Technology, funded by the National Aeronautics and Space Administration and the National Science Foundation.  In this work the Digitized Sky Survey (DSS) and the NASA/IPAC Extragalactic Database (NED) have been used. \'AB acknowledges support provided by NASA through Einstein Postdoctoral Fellowship grant number PF1-120081 awarded by the Chandra X-ray Center, which is operated by the Smithsonian Astrophysical Observatory for NASA under contract NAS8-03060. WF and CJ acknowledge support from the Smithsonian Institution. 
\end{small}

\end{document}